\documentclass[10pt]{aastex631} 
\usepackage{amsmath}
\usepackage{amssymb}
\usepackage{graphicx,float}
\usepackage[utf8]{inputenc}
\usepackage{booktabs}
\usepackage{array,multirow}
\usepackage{gensymb}
\usepackage{tabularray}
\usepackage{longtable}
\begin{document}

\title{Characterizing Gamma-Radio Delayed Flaring Activity from Blazars}

\author[0000-0003-3782-0128]{Alina Kochocki}
\affiliation{\textit{Department of Physics and Astronomy, MSU, East Lansing, MI 48823, USA}}

\author[0000-0003-2769-3591]{Emma Kun}
\affiliation{Department of Astronomy, Institute of Physics and Astronomy, ELTE E\"{o}tv\"{o}s Lor\'{a}nd University, P\'{a}zm\'{a}ny P\'{e}ter s\'{e}t\'{a}ny 1/A, H-1117 Budapest, Hungary}
\affiliation{Konkoly Observatory, HUN-REN Research Centre for Astronomy and Earth Sciences, Konkoly Thege Miklós \'ut 15-17, H-1121 Budapest, Hungary}
\affiliation{CSFK, MTA Centre of Excellence, Konkoly Thege Miklós \'ut 15-17, H-1121 Budapest, Hungary} 

\author[0009-0007-2644-5955]{Sam Hori}
\affiliation{Dept. of Physics and Wisconsin IceCube Particle Astrophysics Center, University of Wisconsin—Madison, Madison, WI 53706, USA}

\begin{abstract}
Flaring activity from the jets of active galactic nuclei has been studied for several decades, closely related to the loading and evolution of the jet. In this work, we focus on the sub-hundred parsec jet region, well traced by non-thermal radio and gamma-ray emission. Only in recent years have light curves capturing the decade-long behavior of such sources become available for a large ensemble of objects. While previous studies have focused on a direct correlation or few-month lag between gamma-ray and radio activity, recent neutrino-bright blazars observed by the IceCube Neutrino Observatory present multi-year delays between initial gamma-ray activity and subsequent radio flares. In this work, we search for similar-timescale correlations between Fermi-LAT gamma-ray data and RATAN-600 radio data from $\sim$100 blazars. We consider two gamma-ray bands, 100 MeV–1 GeV and 1 GeV–500 GeV, as well as the integral band, to compare correlations between potential opaque and unabsorbed regions of the jet. Gaussian process modeling is used for smooth light curve function prediction. We also analyze morphological AGN core data from the MOJAVE survey, forming a sub-selection to better illustrate potential dependence on location. In the broader selection, several sources exhibit delayed flares on the order of 1-3 years. In the stacked analysis, we find the highest correlation for a radio delay on the order of 180 days. The stacked correlation resulting from the MOJAVE sub-selection corresponds to a slightly smaller time lag. Delayed radio flares or extended radio emission appear to be notable features within the general blazar population. 
\end{abstract}

\section{Introduction}

\subsection{Blazar Jet Structure and Activity}

Blazars, active galactic nuclei (AGN) with jets oriented within $\sim$15 degrees of earth, are especially bright due to their relativistically beamed emission. This orientation allows for observation and characterization of their inner-most parsecs. While the longer, extended jet is typically associated with galaxy interaction, these shorter scales allow for study of material loading at the jet base, particle acceleration, and potential radiative or neutrino production \citep{hovatta2020relativisticjetsblazars}. 

Typical models of blazar jet emission assume a compact, collimated jet base where freshly accelerated hadrons and leptons are additionally boosted \citep{Blandford:1977ds}. A strong magnetic field is expected closest to the black hole, decreasing at larger distances. As leptons are accelerated in these magnetic fields, some portion of this energy is dissipated at radio–X-ray energies through synchrotron radiative emission. At higher gamma-ray energies, photons from synchrotron or external photons scatter off the leptonic population through the inverse-Compton process. The distance of material loading, magnetic field strength and material density, largely sets the relative intensity between gamma-ray and radio emission \citep{blum_gould, Cerruti_2020, Petropoulou_2014}.

Several observational effects may impact the observed gamma-ray and radio intensities. As the jet base is smallest in volume, produced photons from synchrotron emission may become self-absorbed, limiting the observable radiation \citep{Potter_2012, Potter_2013, Potter_2015}. Additionally, the inner-parsec is commonly associated with the broad line region, a field of UV photons that may partially absorb gamma-ray emission from certain energy bands \citep{Araudo_2010, Liu_bai}. 

Previous modeling work has proposed quasi-simultaneous flares between gamma-ray and radio data, with potential few-month lags \citep{Boula_2018, Wang_2022}. A number of studies have searched for such correlations, assuming this rough time period as a prior, and found some evidence for this behavior in stacked populations of blazars \citep{Kramarenko_2021, Fuhrmann_2014}. As the exact loading or injection point determines the relative radio or gamma-ray intensity, some sources may present weak or non-detectable flares, adding some complication. 

\subsection{Neutrino-Associated Blazars}

In recent studies performed by the IceCube Neutrino Observatory, several blazars with similar spectral and flaring features have emerged as likely neutrino-bright sources. Specifically, the objects TXS 0506+056, PKS 1424+240 and GB6 J1542+6129 have been associated with local (pre-trial) $\sim$3$\sigma$ significances in a time-integrated analysis \citep{Aartsen_2020}. We list properties of these sources in Table \ref{tab:icecube_srcs}. All sources present similar synchrotron spectral characteristics, with the distribution peaks of the former two cases in the optical to UV range \citep{Padovani_2019, Padovani_2022}. These two sources also show unique flaring signatures where an initial gamma-ray flare coincides with a multi-year delayed radio flare \citep{Kun_2018, 2023Symm...15..270K}. With either source, this previous work has also localized the radio flare to lie within several parsecs of the core region, as opposed to substantially longer distances along the jet. The initial gamma-ray flare for TXS 0506+056 corresponds to the period that the high-energy IceCube alert event, IceCube-170922A, was observed. In both cases, the slow rise of the radio flare begins around the time of the gamma-ray flare, peaking only years later. This suggests a causal relation between an initial jet loading (gamma-ray activity) and later radio activity. While this could be related to observational effects, this may also be caused by the interaction of multiple particle populations or zones. As an example, the propagating material of the jet could interact with older electron populations further along the jet stream, causing a delayed increase in synchrotron radiation. 

\begin{table}[h]
    \centering
    \begin{tabular}{c|c|c|c} \hline 
         Source Name & R.A. [$^\circ$] & Dec. [$^\circ$] & Redshift $(z)$ \\ \hline
         TXS 0506+056 & 77.35 & 5.69 & 0.34  \\
         PKS 1424+240 & 216.76 & 23.80 & 0.61 \\
         GB6 J1542+6129 & 235.75 & 61.50 & $0.34 \leq z \leq 1.76$ \\ \hline
    \end{tabular}
    \caption{Selected, significant IceCube blazars. We provide the right ascension (R.A.) and declination (Dec.) in degrees for the three blazars historically associated with potential neutrino excesses. The redshift of each source is also provided \citep{txs_redshift, Sahu_2024, Padovani_2022}. Notably, all blazars have been traditionally characterized as BL Lac objects, but were more recently found to show substantial evidence for a broad line region \citep{Padovani_2019, Padovani_2022}.   }
    \label{tab:icecube_srcs}
\end{table}

\begin{figure}
\centering

\includegraphics[width=0.85\textwidth]{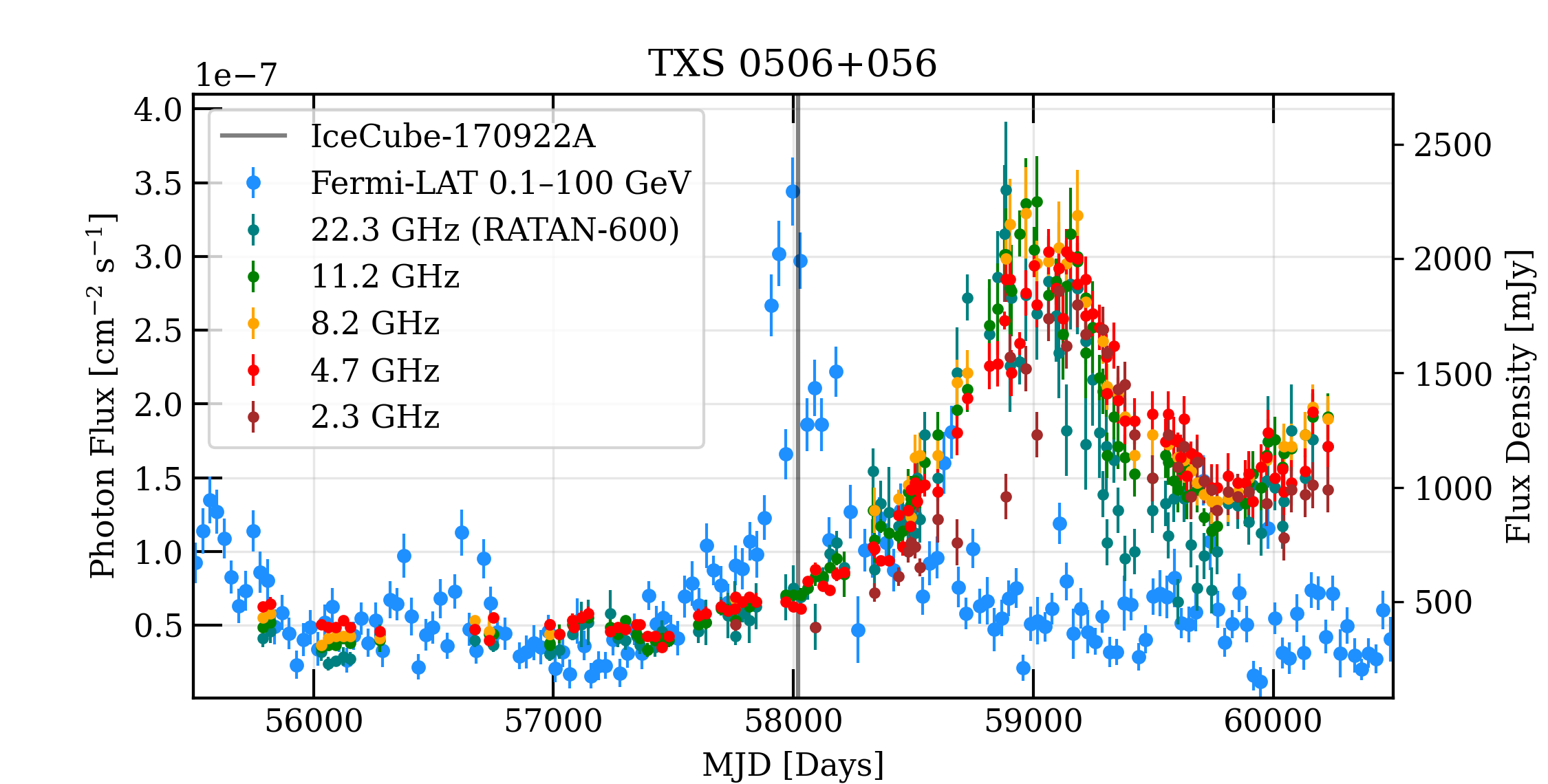}
\vspace{0.3cm}

\includegraphics[width=0.85\textwidth]{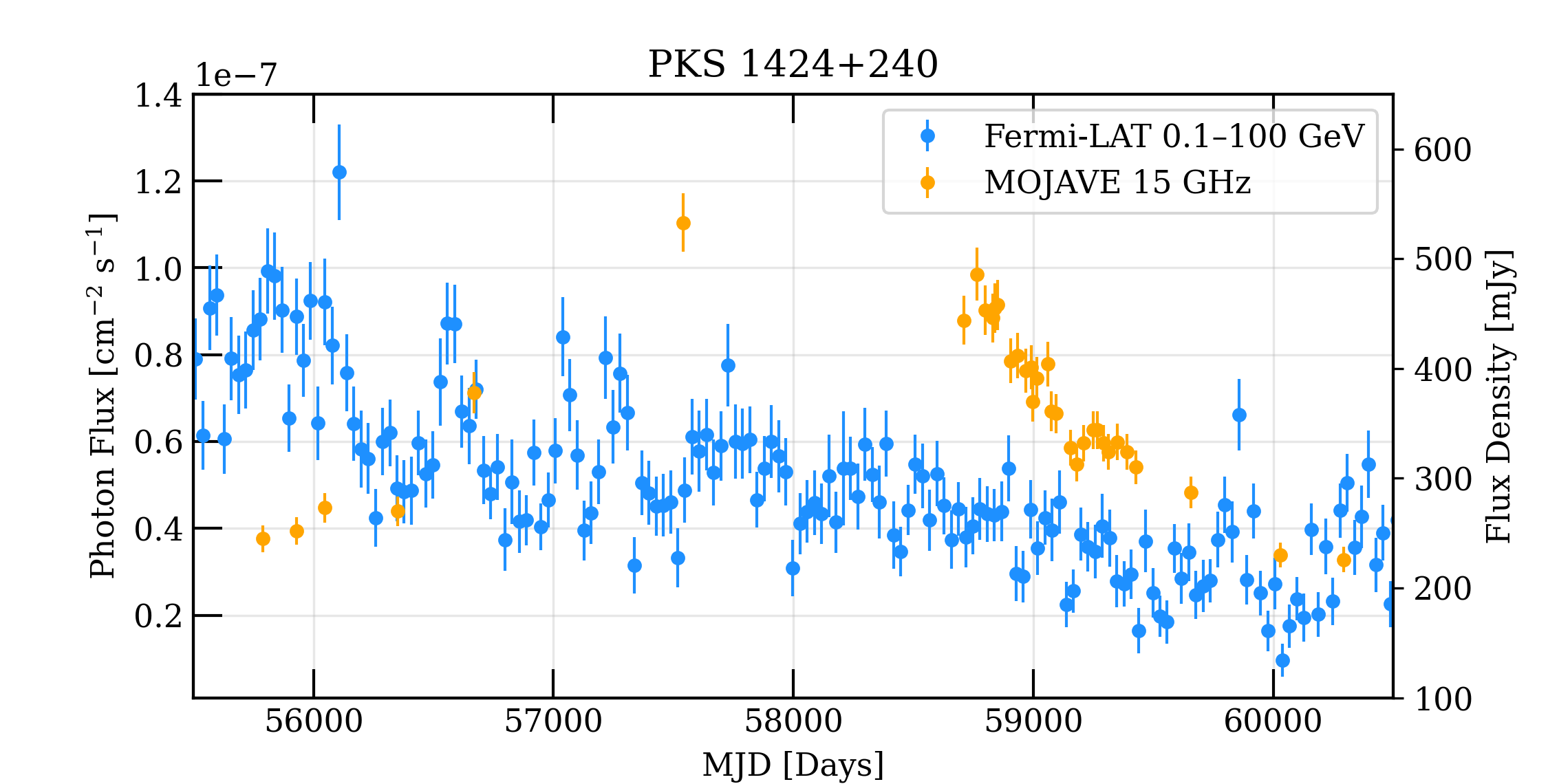}

\caption{Light curves for TXS 0506+056 and PKS 1424+240. In the top plot, we present available RATAN-600 light curves at several frequencies as well as Fermi-LAT data for TXS 0506+056 \citep{Sotnikova_2022, fermi_public}. In the bottom plot, we provide MOJAVE and Fermi-LAT observations for PKS 1424+240 \citep{2018ApJS..234...12L, fermi_public}. The left axis corresponds to the photon flux of the gamma-ray data, while the right axis represents the flux density of the radio data. In either case, a substantial radio flare follows a period of gamma-ray activity on a delay timescale of a few years. The radio intensity appears to begin to increase after the end of the gamma-ray-active period. }
\label{fig:txs_pks_curves}

\end{figure}

In Figure \ref{fig:txs_pks_curves}, we provide a comparison of archival gamma-ray data and radio data for the two sources, TXS 0506+056 and PKS 1424+240. While the third source, GB6 J1542+6129, shows some potential for a delay in radio emission, these light curves are more variable, potentially presenting emission from multiple zones.

\subsection{A Search for Delayed Radio Flares}

In this work, we investigate the general existence of this flaring behavior in a wider selection of $\sim$100 blazars. We focus on a correlation between Fermi-LAT gamma-ray data and RATAN-600 radio data with a -0.5 to 3.5 year relative lag. We consider correlations with two separate gamma-ray bands, 100 MeV–1 GeV and 1 GeV–500 GeV, as well as the integral band, to better compare the significance of any correlation between potentially gamma-ray opaque and transparent regions. We also utilize MOJAVE \citep{2018ApJS..234...12L} morphological radio data to form an additional sub-selection of core-flaring blazars, where we are more sensitive to particle injection localized to the jet base.

Our source selection is based on the availability of high-cadence, long–duration RATAN–600 blazar light curves exhibiting some evidence of variability. We use this selection to perform new Fermi–LAT light curve analyses at the relevant source locations for the described energy bands. While these data form the basis of our correlation analyses, we further analyze available MOJAVE morphological data to form an additional sub-selection of core–flaring blazars. We consider time-dependent correlations for individual sources as well as for aggregate, stacked blazar populations. 

While this analysis is largely motivated by the phenomenology of neutrino-flaring blazars, we generally expect our study to better illustrate the multi-zone morphology of sub-hundred parsec AGN jets. If a substantial delayed correlation was observed, this might indicate a more complex density of particle populations along the jet stream. 

\section{Source Selection and Multiwavelength Surveys}

\subsection{RATAN-600 Radio Light-Curve Data}

The RATAN–600 telescope is a centimeter–meter wavelength telescope located at the Special Astrophysical Observatory (SAO) in Russia. The telescope features a 576-meter diameter reflective ring, capable of adjusting to a range of declinations. A secondary reflective system also houses multiple receivers spanning roughly 1–30 GHz in frequency. The telescope has operated for several decades, providing exhaustive, long–term monitoring of radio-bright blazars at multiple wavelengths and declinations. 

In this work, we consider available public blazar light curves for a range of observing wavelengths \citep{Sotnikova_2022}. As observing cadence varies substantially for different wavelengths between sources, we consider 7.7, 8.2, 11.2, 14.4, and 21.7 GHz data. While the level of self-absorption between these different wavelengths may very slightly, if a stacked set of light curves is tested, we generally expect this variation to be a sub-dominant effect in any result. The variation in absorption between these frequencies may lead to a shifted lag on the order of a few months as opposed to years. 

We consider three selection criteria for a source's set of light curves, $\ell_{f}$, of frequency, $f_{j} \in \{7.7, 8.2, 11.2, 14.4, 21.7\}$ GHz, where $j$ indexes the available frequencies. Given $N$ times, $t_{j,i}$, corresponding flux densities, $S_{j,i}$, and flux density uncertainties, $\sigma_{j,i}$, we select such that, 
\begin{enumerate}
    \item $ \textrm{max}_{i}( t_{j, i} ) - \textrm{min}_{i}(t_{j, i} ) > 8$ years. Given the $j$th light curve for a specific flux density, this ensures a long enough observing period to examine multiple phases of radio activity. 
    \item Given all adjacent pairs of observations, $t_{j, i + 1} - t_{j, i} < 1$ year. This ensures timing resolution at least on the order of one year for our light curves. 
    \item We choose the final light curve for a source, $\ell_{f_{ \Delta \textrm{min}}}$, such that $f_{ \Delta \textrm{min}}$ corresponds to $ t_{\Delta \textrm{min}} =  \textrm{min}_{j} ( \textrm{max}_{i}( t_{j, i + 1} - t_{j, i} ) )  $. The frequency selected is the light curve whose sampling gap, or largest period between subsequent measurements, is minimal across frequencies. 
\end{enumerate}
If no light curve of any frequency satisfies these selection criteria, the source is not considered further. The result is 231 objects. We also make a selection based on the fractional excess variance of each light curve, and whether the RATAN object location is spatially coincident with a known Fermi–LAT source.
\begin{enumerate}
    \item We define the mean flux, $\bar{S}$, sample variance of the fluxes, $s^{2}$, and measurement error, $\sigma^{2}$, as the following, 
    \begin{align}
        \bar{S} &= \dfrac{1}{N} \sum^{N}_{i = 1}S_{i}, \\
        s^{2} &= \dfrac{1}{N - 1} \sum^{N}_{i = 1}(S_{i} - \bar{S})^{2}, \\
        \sigma^{2} &= \dfrac{1}{N} \sum^{N}_{i = 1} \sigma_{i}^{2}.
    \end{align}
    The fractional excess variance is then, 
    \begin{equation}
        F_{\textrm{var.}} = \dfrac{ \sqrt{ | s^{2} - \sigma^{2} | } }{ \bar{S} }.
    \end{equation}
    We select sources such that $F_{\textrm{var.}}(\ell_{f_{ \Delta \textrm{min}}}) > 0.1$. Each chosen light curve then exhibits some evidence of variability or flaring over the observed period, with respect to the flux density measurement uncertainties. 
    \item We also compare the location of each source to objects within the Fermi 4LAC catalog. If the closest object exists with an angular separation less than 0.008 degrees, we retain the object. 
\end{enumerate}
We find that 104 sources pass these selection criteria.

\subsection{Fermi-LAT Gamma-ray Light-Curve Data}

The Fermi Large Area Telescope (Fermi–LAT) is a satellite observatory capable of indirect gamma-ray detection \citep{2009ApJ...697.1071A}. High-energy photons that interact within the silicon–tungsten tracking-calorimeter system produce electron–positron pairs. The tracker system is capable of reconstructing event angular uncertainty with sub-degree resolution above 1 GeV. A calorimeter measurement determines energy for the particle interaction. In general, Fermi–LAT provides sensitivity to an energy range between 20 MeV and 300 GeV. As the survey view is partially obstructed by the Earth, full-sky cadence is achieved only every several hours. 

In this work, we perform binned likelihood analyses with roughly fifteen years of Fermi–LAT data for our selection of $\sim$100 blazars. The data selection spans the period between December 11, 2009 and November 5, 2024. We consider three energy bands for each source, 100 MeV–1 GeV, 1 GeV–500 GeV, and 100 MeV–500 GeV. As certain sources are relatively dim gamma-ray objects, we consider a conservative time binning scheme with a resolution of 22 bins over the observing period. In each bin, we fit the spectral parameterization suggested by the LAT 14-year source catalog, generally a power-law energy spectrum with free index and normalization.

Following the standard data analysis procedure recommended by Fermi, we download the relevant observational data, spacecraft data and supplemental data products. We first filter our data. We assume an event class of `128' and event type of `3' to select only events with a high probability of being a photon interaction that may be either front or back converting. This is the standard event quality cut used for Fermi point sources. We use a maximum zenith cut of 90 degrees with our event filter to remove photons from Earth's limb. Events with reconstructed energies between 100 MeV and 500 GeV are used for the general analysis. 

Observed counts, the livetime and exposure map are determined as a function of space and energy. We use instrument response functions based on the first eight years of observation, Pass 8 P9R3. Standard models provided by Fermi are used for the expected, isotropic extragalactic emission and galactic emission. Basic settings follow previous, generic Fermi light curve tutorials. We use the Python package, LATSourceModel, to construct a source list referencing the LAT 14-year source catalog. Sources within 5 degrees of the target location are fit with free spectral parameters if above a 5$\sigma$ significance threshold. Flux expectations are generated for the selection of sources. Finally, we construct our binned observation and analysis objects based on these results, and perform our binned likelihood analyses for the described energy ranges. 

The test statistic ($TS$) represents the relative likelihood ratio between the best-fit result, $L$, and null result, $L_{0}$, such that $TS = \textrm{2 ln}(L/L_{0})$. With each binned measurement, if a test statistic value $> 4$ was found, the symmetrized $68 \%$ confidence interval was determined. Otherwise, a 95$\%$ upper limit on the photon flux was instead placed. Given that a faint source may be dominated by upper limits, we further consider only gamma-ray light curves with at least 7 significant measurements (non upper limit). We used the determined photon fluxes within our correlation analyses.


\subsection{MOJAVE Radio Morphology Data}

\subsection{VLBI core flux extraction for RATAN blazars with MOJAVE coverage}

To quantify what fraction of the parsec-scale 15\,GHz radio emission originates from the compact VLBI core, we analyzed those RATAN-monitored blazars that have Very Long Baseline Array (VLBA) coverage within the MOJAVE program at 15\,GHz \citep{2018ApJS..234...12L}. The calibrated complex visibilities were obtained from the public MOJAVE archive. The data reduction and model-fitting strategy follow the procedure described in \citet{Kun2014,Kun2015}.

For each observing epoch, the calibrated visibility data were imaged and self-calibrated in \texttt{DIFMAP} \citep{1997ASPC..125...77S}. After standard phase and amplitude self-calibration cycles, we produced naturally weighted CLEAN maps with typical image sizes of $1024\times1024$ pixels and a pixel scale of 0.04\,mas. The residual rms noise from the map and the restoring beam parameters were recorded for each epoch. The brightness distribution was parameterized by fitting circular Gaussian components directly to the visibility data. The first component of the model, located at the brightness peak and fixed to the center of the map, was identified as the VLBI core. Additional components were iteratively added to describe downstream jet features until no significant residual structure remained above a predefined signal-to-noise threshold ($\sim 6 \sigma$). This procedure yields, for each epoch, the flux density, size, and position of all fitted components.

The VLBI core flux density $S_{\mathrm{core}}$ was extracted as the flux of the first Gaussian component in the final model. The associated uncertainty was estimated from the image rms noise $\sigma_{\mathrm{rms}}$, the restoring beam size, and the fit component size, accounting for both thermal noise and the signal-to-noise ratio of the component. For each source, this resulted in a time series of 15\,GHz core flux densities at the MOJAVE observing epochs.

 The core dominance parameter was then defined as
\begin{equation}
    f_{\mathrm{core}} = \frac{S_{\mathrm{core}}^{\mathrm{VLBI}}}{S_{\mathrm{tot}}^{\mathrm{VLBI}}},
\end{equation}
where $S_{\mathrm{tot}}^{\mathrm{VLBI}}$ is the quasi-simultaneous single-dish flux density at the closest available frequency to 15\,GHz.

This approach allows us to quantify the fraction of the total radio emission that arises from the unresolved parsec-scale core, and to investigate whether major flares in the RATAN light curves are dominated by the compact core or by more extended jet components. The analysis provides a direct link between single-dish variability and structural changes on milliarcsecond scales.

We defined a conservative VLBI core-dominated subsample by requiring both a high maximum core fraction, $\max(S_{\mathrm{core}}/S_{\mathrm{tot}}) > 0.7$, and a strong correlation between the core-flux and total-flux VLBI light curves, quantified by a Pearson correlation coefficient $r > 0.8$. These thresholds were chosen pragmatically to isolate sources whose parsec-scale 15\,GHz emission is clearly dominated by the compact core, rather than to define a unique physical boundary. This selection yielded 51 sources.

\section{Correlation and Gaussian Process Modeling}

Our correlation analysis uses Gaussian process modeling for smooth light curve construction. As our selection includes several gamma-ray-dim sources, several Fermi–LAT light curves result in only $\sim$10 high-significance data values. This low number of measurements is not ideal for common correlation schemes (e.g. a z–transformed discrete correlation function). Instead, we use a Gaussian process model to infer a smooth underlying light curve consistent with the observed data and measurement uncertainties.

We use the Python package, celerite2, for Gaussian process (GP) modeling \citep{celerite2}. We assume each light curve can be described by the covariance of a stochastically driven, damped harmonic oscillator. This kernel allows for modeling of generally quasi-periodic, damped or noisy data. We can consider the form, 
\begin{equation}
    \ddot{x}(t) + \dfrac{\omega_{0}}{Q} \dot{x}(t) + \omega_{0}^{2}x(t) = \sqrt{S_{0}} \xi(t).
\end{equation}
Here, $\omega_{0}$ determines the angular frequency, $Q$ determines the relative damping rate, and $S_{0}$ is the amplitude of stochastic driving energy. $\xi(t)$ is the unit-variance Gaussian driving function or white noise. We can also relate these values to the parameterization $\{\rho, \sigma, Q\}$, 
\begin{align}
    \rho &= \dfrac{2 Q}{\omega_{0}}, \\
    \sigma^{2} &= \dfrac{S_{0}}{2 \omega_{0} Q}. 
\end{align}
Here, $\rho$ acts as the characteristic damping timescale and $\sigma$ is the root mean square (RMS) amplitude of the oscillator. Gaussian process modeling assumes our input data or time series is jointly Gaussian with covariance determined by our kernel assumption. We define a model object as a function of $\rho, \sigma$, and $Q$, as well as the data uncertainties and times. We are then able to evaluate the likelihood of our observed data points given this constructed model. 

In each case, we scale the input flux (density) values and flux (density) uncertainties relative to the mean flux (density). We assume $Q = 0.5$, $\sigma < 1$, and $\rho < 3000$ days. We optimize the kernel hyperparameters, $\sigma$ and $\rho$ by maximizing the GP log-likelihood. Finally, we compute the GP posterior predictive distribution at the desired evaluation times. This determines the posterior mean and covariance as our predicted function or smooth light curve and uncertainties. 

\begin{figure*}
\centering

\gridline{
  \fig{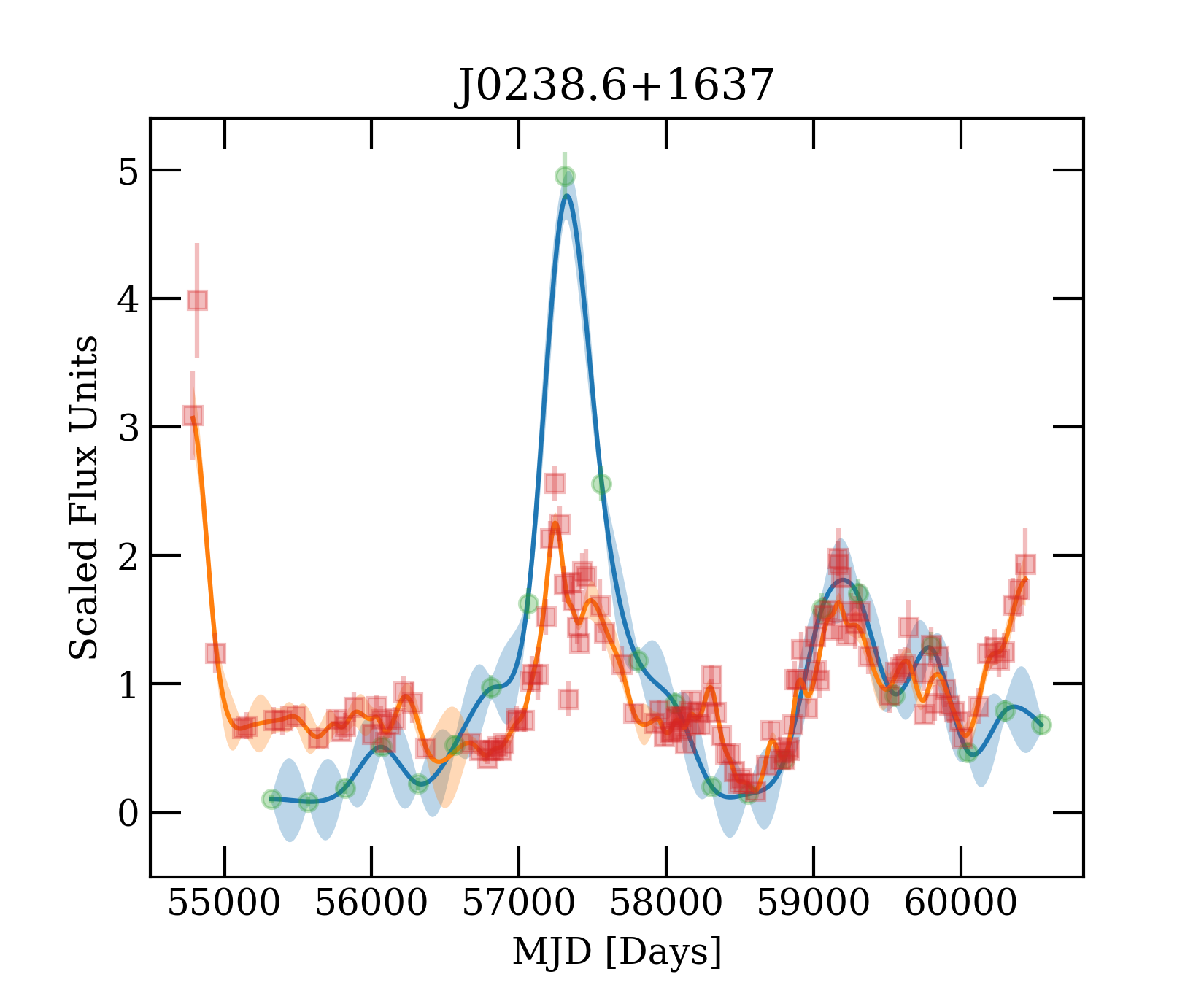}{0.5\textwidth}{}
  \fig{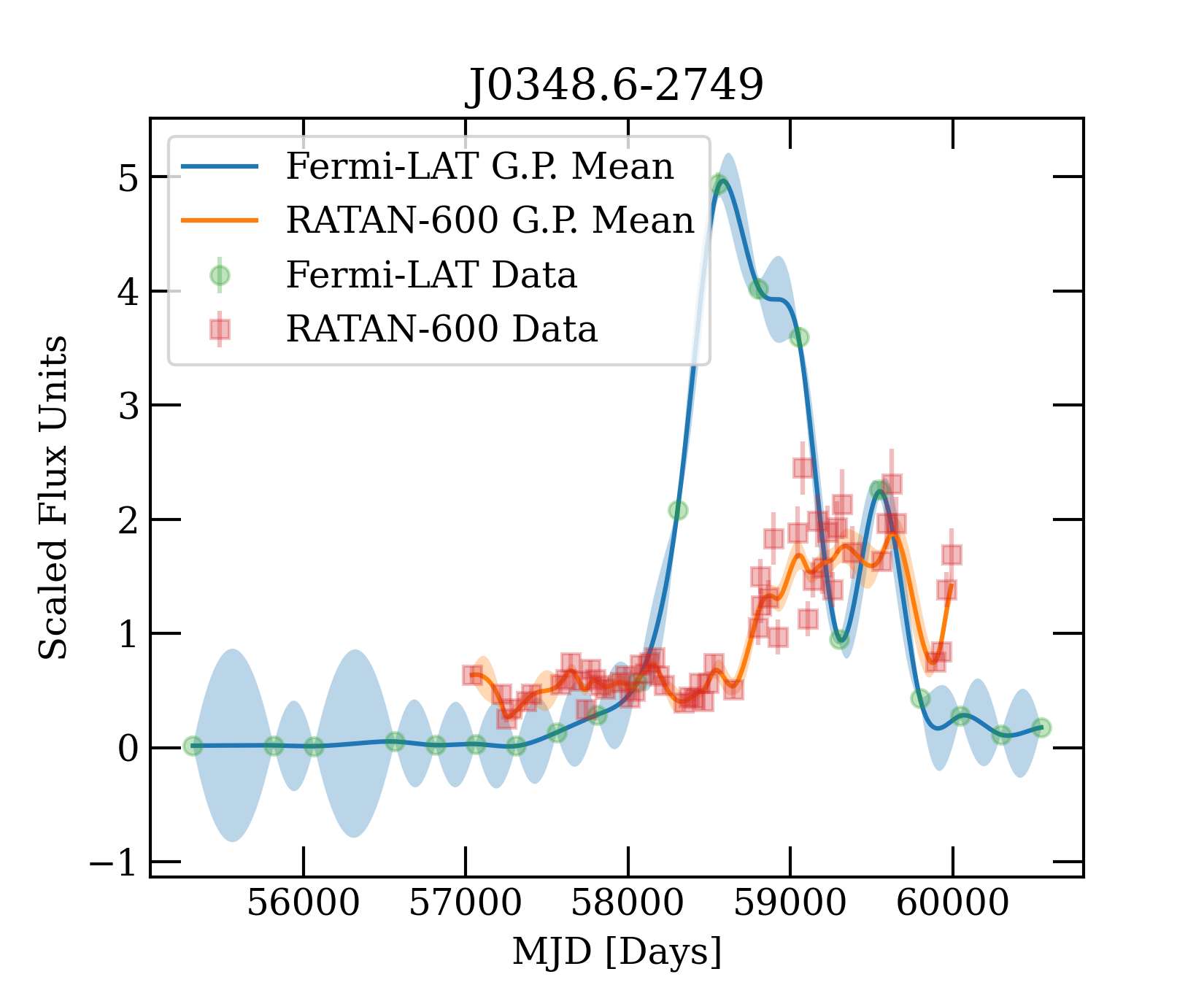}{0.5\textwidth}{}
}
\vspace{-3.0em}
\gridline{
  \fig{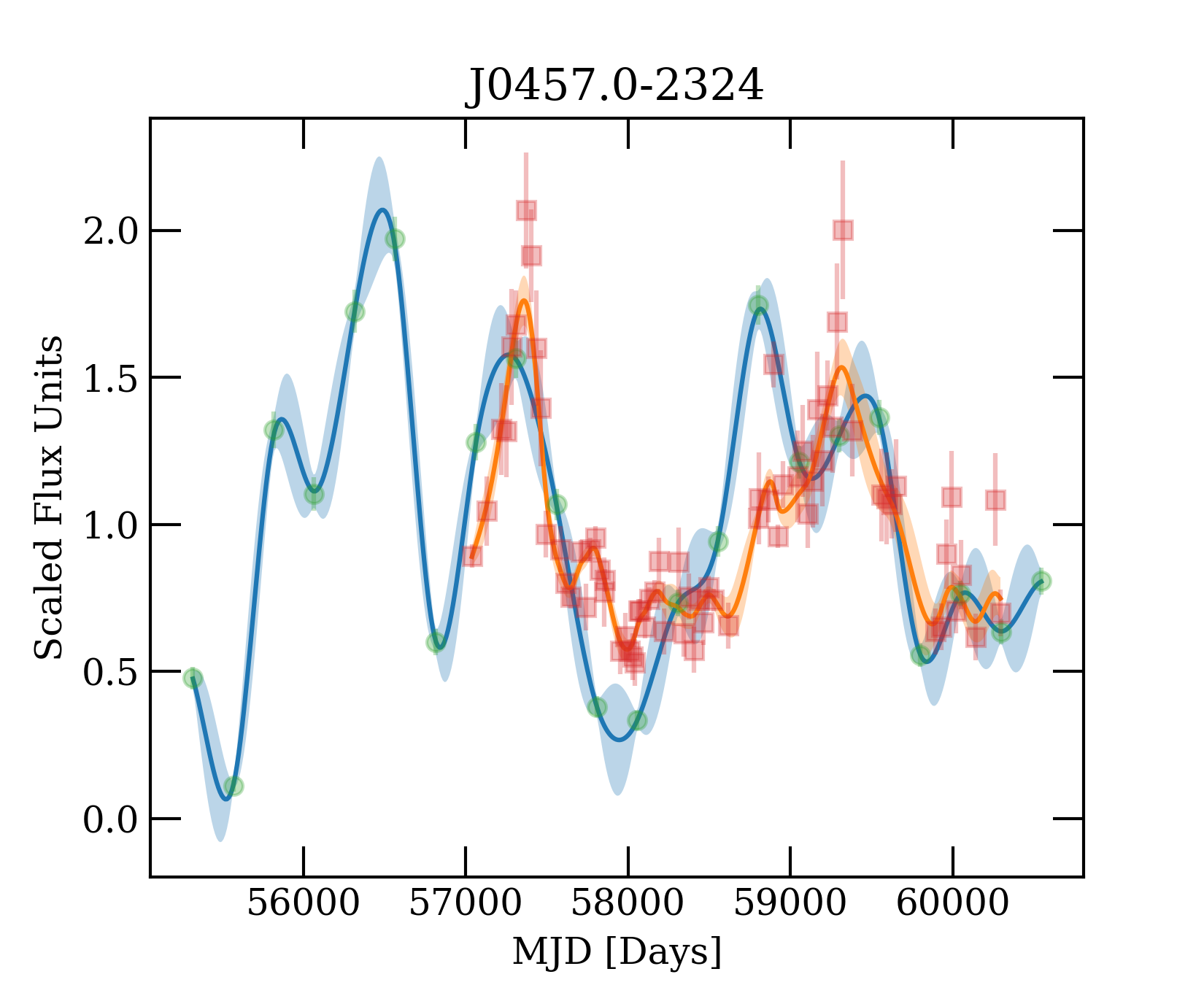}{0.5\textwidth}{}
  \fig{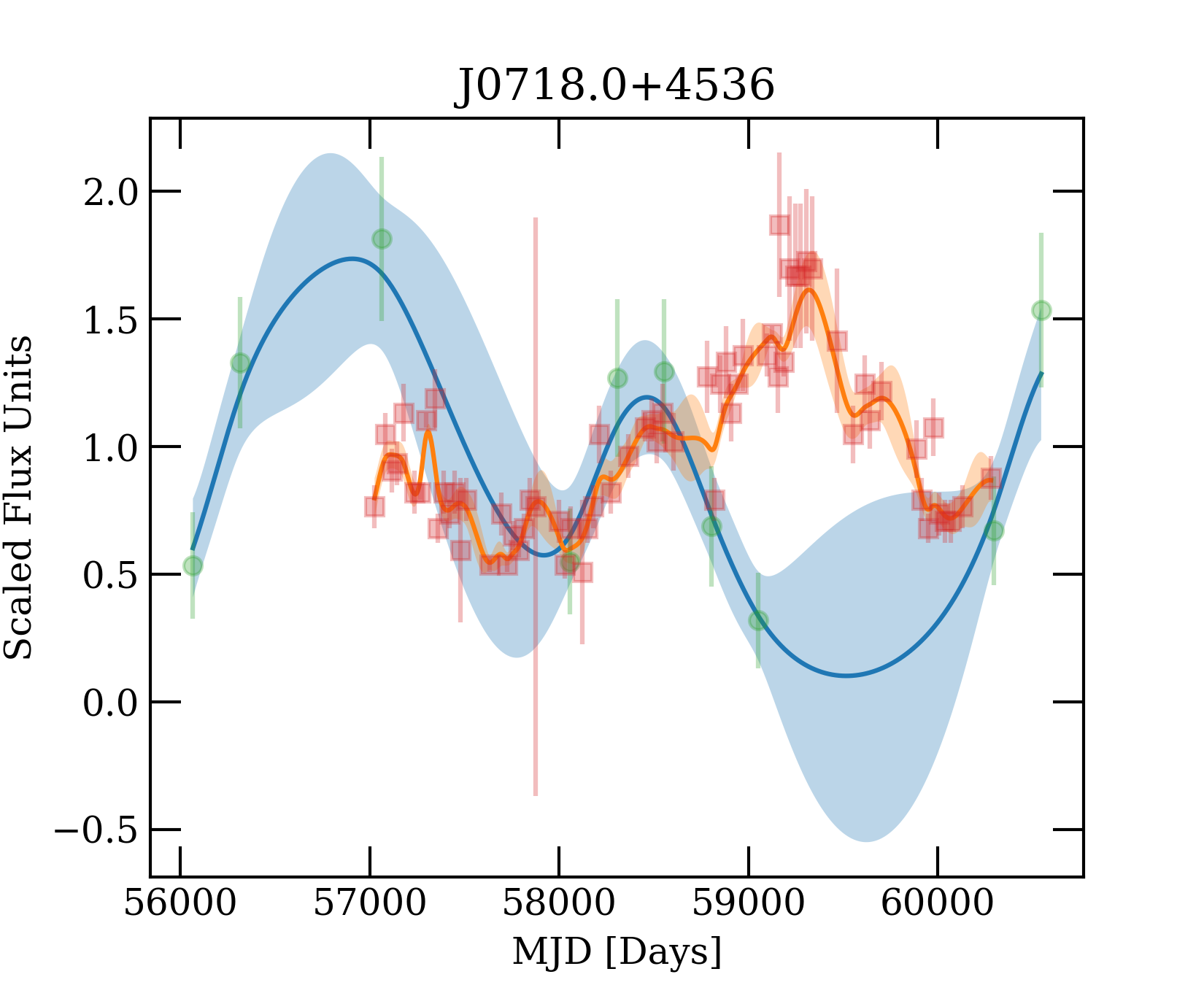}{0.5\textwidth}{}
}
\vspace{-1.5em}
\caption{Representative light curves and uncertainties predicted through Gaussian process modeling. Here, we provide several light curve examples produced from Gaussian process modeling along with the underlying data. The name of the source is provided as the title in each case. The predicted means are plotted, as well as the $1\sigma$ posterior predictive uncertainties as shaded regions. The Fermi photon flux high-energy band is used for all examples. This selection is chosen to be generally representative of the light curve measurement quality observed within the ensemble of $\sim$100 sources. Upper limits are not used in our Gaussian process modeling, and the sparsity of pictured, significant measurements generally represents the number of measurements made per source.}
\label{fig:interp_lc_examples}

\end{figure*}

We determine the predicted light curve for the radio and gamma-ray photon data of each source and Fermi–LAT analysis configuration. We provide examples for four sources in Figure \ref{fig:interp_lc_examples}. We then test a range of sampled lag values, $T \in [-0.5, 3.5]$ years. If the light curve overlap is insufficient to consider the entire lag range, we discard the source. Given a lag value, our smooth gamma-ray light curve is shifted forward in time. The shifted gamma-ray curve is evaluated at the radio observation times for comparison. We then determine the Pearson product-moment correlation coefficient, between the two time series, $R \in [-1,1]$, for the tested lag. We are able to determine the lag value, $T_{\textrm{best}}$, in this range with the highest correlation coefficient. 

To estimate an uncertainty range for this lag value, we generate 30 realizations of each data time series with additive perturbations. These perturbations are sampled from a Gaussian distribution with standard deviation determined from the original measurement uncertainty. We again determine the predicted smooth light curve for the synthetic data given our highest-likelihood Gaussian process model parameterization. We scan the same range of lag values and determine the maximal correlation coefficient and corresponding lag. Finally, we take the standard deviation of this selected lag distribution as the lag uncertainty, $\Delta T_{\textrm{best}}$. 

In this study, it is also relevant to consider the strength of temporal correlation within a stacked population. If we are generally insensitive to a potential significant correlation with an individual source, but the underlying trend is present within a larger population, considering this larger statistical population may allow for detection. With $M$ sources, we construct stacked time series by appending each smooth, predicted curve, and properly incrementing each time value. This is done in such a way that flux measurements for a given source retain their time separation and ordering, and the time delay between sources within the stacked time series is sufficient to prevent overlap of any time series between different sources for the test range of lags. The lag value corresponding to the highest correlation coefficient is determined. We again use the perturbed synthetic data realizations to estimate the lag uncertainty. 

Given either an individual source or stacked analysis, we are also able to estimate the result significance. In the case of of the individual source analysis, we perform additional correlations between the radio light curve and mismatched gamma-ray light curves. For a larger statistical sample, we also shift the selected gamma-ray light curve in time by a random value within the range of a year. These trials allow us to construct central $68\%$ and $95\%$ containment regions for our correlation coefficients as a function of lag. In the case of the stacked light curve, we again select a randomized order of gamma-ray light curves to construct the gamma-ray time series. 
\newpage

\section{Individual-Source Correlations}

We present the correlation results for our individual source studies at low and high gamma-ray energy bands in Table \ref{tab:delayed_corr_results}. Additional results for the integral energy band are listed with Figure \ref{tab:delayed_corr_results_int} within the Appendix. Several sources show some evidence for best-fit correlations within a delayed window of 0.5 to 3.5 years.

\begin{longtable}{lccccccccc}
\caption{Correlation results for a radio-delayed flare. Given our search window of -0.5 to 3.5 years (-182.5 to 1277.5 days), we provide the tested lag and uncertainty, $T_{\textrm{delay}}$ and $\Delta T_{\textrm{delay}}$ in units of days, corresponding to the highest correlation coefficient $R_{\textrm{delay}}$. Results for both the low-energy and high-energy gamma-ray band are provided. If one of the energy-band light curves did not pass our selection criteria for a given source, the results are not listed. } \\
\hline
 Source Name & R.A. [$\degree$] & Dec. [$\degree$] & $T_{\textrm{delay, low}}$ &  $\Delta T_{\textrm{delay, low}}$ &  $R_{\textrm{delay, low}}$  & $T_{\textrm{delay, high}}$ &  $\Delta T_{\textrm{delay, high}}$ &  $R_{\textrm{delay, high}}$ \\ \hline
\endfirsthead

\hline
 Source Name & R.A. [$\degree$] & Dec. [$\degree$] & $T_{\textrm{delay, low}}$ &  $\Delta T_{\textrm{delay, low}}$ &  $R_{\textrm{delay, low}}$  & $T_{\textrm{delay, high}}$ &  $\Delta T_{\textrm{delay, high}}$ &  $R_{\textrm{delay, high}}$ \\ \hline
\endhead

J0018.8+2611 & 4.91 & 26.05 & 725.12 & 120.0 & 0.8 & -- & -- & -- \\ 
J0112.8+3208 & 18.21 & 32.14 & 625.05 & 163.87 & 0.13 & 578.02 & 130.49 & 0.22 \\ 
J0118.9-2141 & 19.74 & -21.69 & 619.05 & 16.35 & 0.68 & 252.8 & 158.54 & 0.74 \\ 
J0132.7-1654 & 23.18 & -16.91 & 534.99 & 15.81 & 0.81 & 523.98 & 13.63 & 0.71 \\ 
J0144.6+2705 & 26.14 & 27.08 & 311.84 & 187.68 & 0.36 & 144.72 & 111.83 & 0.6 \\ 
J0152.2+2206 & 28.07 & 22.12 & 319.84 & 383.57 & 0.6 & -182.5 & 264.96 & 0.45 \\ 
J0238.6+1637 & 39.66 & 16.62 & -0.38 & 12.46 & 0.79 & -0.38 & 10.87 & 0.78 \\ 
J0239.7+0415 & 39.96 & 4.27 & 685.09 & 102.74 & 0.85 & 611.04 & 167.06 & 0.84 \\ 
J0312.8+0134 & 48.18 & 1.55 & -- & -- & -- & 319.84 & 501.43 & 0.5 \\ 
J0336.4+3224 & 54.12 & 32.31 & 664.08 & 375.03 & 0.93 & -- & -- & -- \\ 
J0343.4+3621 & 55.87 & 36.37 & -- & -- & -- & 1229.47 & 482.01 & 0.68 \\ 
J0348.6-2749 & 57.16 & -27.82 & 578.02 & 22.65 & 0.96 & 588.03 & 20.48 & 0.95 \\ 
J0405.6-1308 & 61.39 & -13.14 & 141.72 & 58.35 & 0.75 & 253.8 & 96.73 & 0.88 \\ 
J0416.5-1852 & 64.15 & -18.85 & 50.66 & 376.33 & 0.79 & -182.5 & 714.01 & 0.4 \\ 
J0423.3-0120 & 65.81 & -1.34 & 269.81 & 90.9 & 0.79 & -182.5 & 165.93 & 0.75 \\ 
J0423.9+4150 & 65.98 & 41.83 & -128.46 & 562.18 & 0.65 & -156.48 & 570.1 & 0.59 \\ 
J0434.1-2014 & 68.53 & -20.25 & -- & -- & -- & -182.5 & 635.09 & -0.01 \\ 
J0442.6-0017 & 70.66 & -0.3 & -182.5 & 285.3 & 0.45 & -182.5 & 264.73 & 0.47 \\ 
J0449.1+1121 & 72.28 & 11.36 & 30.65 & 19.16 & 0.91 & 30.65 & 153.51 & 0.83 \\ 
J0453.1-2806 & 73.31 & -28.13 & 910.25 & 53.14 & 0.44 & 79.68 & 211.4 & 0.49 \\ 
J0457.0-2324 & 74.26 & -23.41 & 125.71 & 25.29 & 0.59 & 72.67 & 20.92 & 0.77 \\ 
J0502.5+1340 & 75.64 & 13.64 & -- & -- & -- & -182.5 & 311.7 & 0.95 \\ 
J0510.0+1800 & 77.51 & 18.01 & -65.42 & 11.95 & 0.56 & -56.41 & 11.27 & 0.55 \\ 
J0532.6+0732 & 83.16 & 7.55 & 1159.42 & 491.19 & 0.6 & 1138.4 & 504.63 & 0.5 \\ 
J0533.3+4823 & 83.31 & 48.38 & 383.89 & 25.36 & 0.91 & 349.86 & 103.39 & 0.72 \\ 
J0539.9-2839 & 84.97 & -28.67 & 293.83 & 33.9 & 0.43 & 220.78 & 43.69 & 0.53 \\ 
J0654.4+4514 & 103.6 & 45.24 & -120.46 & 574.37 & 0.65 & 88.69 & 337.78 & 0.56 \\ 
J0656.3-0322 & 104.05 & -3.38 & 1277.5 & 638.35 & 0.05 & 40.65 & 450.3 & 0.34 \\ 
J0659.6-2742 & 104.95 & -27.75 & 450.93 & 257.19 & 0.62 & 214.77 & 211.75 & 0.63 \\ 
J0718.0+4536 & 109.46 & 45.63 & 743.13 & 104.98 & 0.34 & 632.06 & 356.28 & -0.22 \\ 
J0725.2+1425 & 111.32 & 14.42 & 539.99 & 105.75 & 0.27 & 676.09 & 179.75 & 0.21 \\ 
J0748.6+2400 & 117.15 & 24.01 & 753.14 & 149.33 & 0.74 & -182.5 & 235.54 & 0.42 \\ 
J0754.4-1148 & 118.61 & -11.79 & -74.43 & 129.46 & 0.83 & -85.43 & 144.68 & 0.29 \\ 
J0808.2-0751 & 122.06 & -7.85 & 78.68 & 26.95 & 0.86 & 78.68 & 19.21 & 0.83 \\ 
J0809.3+4053 & 122.23 & 40.88 & 735.13 & 99.6 & 0.5 & 599.04 & 256.56 & -0.09 \\ 
J0829.0+1755 & 127.27 & 17.9 & -141.47 & 385.24 & 0.37 & 409.91 & 351.76 & 0.31 \\ 
J0830.8+2410 & 127.72 & 24.18 & 262.81 & 114.63 & 0.5 & 206.77 & 83.09 & 0.61 \\ 
J0850.0+5108 & 132.49 & 51.14 & -84.43 & 355.44 & 0.23 & 98.69 & 162.47 & 0.34 \\ 
J0920.9+4441 & 140.24 & 44.7 & 46.66 & 57.77 & 0.72 & 62.67 & 46.16 & 0.83 \\ 
J0921.6+6216 & 140.4 & 62.26 & -2.38 & 119.62 & 0.64 & 259.8 & 77.15 & 0.66 \\ 
J0923.5+4125 & 140.88 & 41.42 & 81.68 & 20.26 & 0.23 & 948.27 & 315.44 & -0.41 \\ 
J1006.7-2159 & 151.69 & -21.99 & 1045.34 & 207.93 & 0.27 & 594.03 & 176.04 & 0.31 \\ 
J1114.5-0819 & 168.63 & -8.28 & 85.68 & 432.82 & 0.39 & 732.13 & 571.84 & 0.13 \\ 
J1118.6-1235 & 169.57 & -12.55 & 630.06 & 94.38 & 0.87 & 1237.47 & 478.42 & 0.79 \\ 
J1127.0-1857 & 171.77 & -18.95 & 1030.33 & 8.06 & 0.93 & 1029.33 & 11.75 & 0.94 \\ 
J1129.8-1447 & 172.53 & -14.82 & 939.27 & 20.75 & 0.8 & 995.31 & 154.75 & 0.81 \\ 
J1135.7-0427 & 173.99 & -4.47 & 1131.4 & 589.78 & 0.22 & -- & -- & -- \\ 
J1345.8+0706 & 206.45 & 7.11 & 817.18 & 296.4 & 0.25 & 745.14 & 347.01 & 0.26 \\ 
J1357.1+1921 & 209.27 & 19.32 & 809.18 & 361.73 & 0.66 & -- & -- & -- \\ 
J1613.6+3411 & 243.42 & 34.21 & 725.12 & 30.81 & 0.74 & 511.98 & 209.64 & 0.89 \\ 
J1616.7+4107 & 244.27 & 41.11 & 412.91 & 312.17 & 0.46 & -182.5 & 678.96 & -0.16 \\ 
J1631.2+4926 & 247.82 & 49.46 & 964.29 & 377.93 & 0.85 & 1277.5 & 667.89 & -0.02 \\ 
J1635.2+3808 & 248.81 & 38.13 & 85.68 & 6.67 & 0.73 & 89.69 & 8.09 & 0.72 \\ 
J1642.9+3948 & 250.74 & 39.81 & -106.45 & 37.99 & 0.71 & 1063.35 & 405.52 & 0.67 \\ 
J1728.4+0427 & 262.1 & 4.45 & 1277.5 & 370.66 & 0.25 & 1277.5 & 585.81 & 0.22 \\ 
J1733.0-1305 & 263.26 & -13.08 & 1277.5 & 75.94 & 0.4 & 1277.5 & 0.62 & 0.43 \\ 
J1734.3+3858 & 263.58 & 38.96 & 669.08 & 14.25 & 0.45 & 690.1 & 129.14 & 0.37 \\ 
J1740.5+5211 & 265.15 & 52.2 & 276.81 & 20.82 & 0.7 & 224.78 & 25.02 & 0.64 \\ 
J1814.4+2953 & 273.4 & 29.88 & 849.21 & 127.05 & 0.27 & -- & -- & -- \\ 
J2136.2+0032 & 324.16 & 0.7 & 1277.5 & 90.46 & 0.44 & -- & -- & -- \\ 
J2146.4-1528 & 326.59 & -15.43 & 117.71 & 612.94 & 0.88 & -- & -- & -- \\ 
J2147.1+0931 & 326.79 & 9.5 & 1277.5 & 137.62 & 0.21 & 92.69 & 533.38 & 0.28 \\ 
J2158.1-1501 & 329.52 & -15.02 & 309.84 & 275.21 & 0.25 & 392.89 & 347.83 & -0.03 \\ 
J2219.2-0342 & 334.72 & -3.59 & 1277.5 & 548.17 & 0.44 & 254.8 & 96.22 & 0.87 \\ 
J2219.2+1806 & 334.81 & 18.11 & -35.4 & 9.17 & 0.72 & 451.93 & 218.21 & 0.47 \\ 
J2225.6+2120 & 336.41 & 21.3 & 176.75 & 278.68 & 0.48 & -- & -- & -- \\ 
J2229.7-0832 & 337.42 & -8.55 & -13.38 & 12.16 & 0.84 & -31.4 & 35.37 & 0.84 \\ 
J2232.6+1143 & 338.15 & 11.73 & 27.64 & 57.73 & -0.09 & 236.79 & 200.42 & -0.1 \\ 
J2236.3+2828 & 339.09 & 28.48 & 60.67 & 24.52 & 0.56 & 81.68 & 21.45 & 0.51 \\ 
J2243.9+2021 & 340.97 & 20.35 & -66.42 & 413.31 & -0.55 & 449.93 & 123.01 & 0.57 \\ 
J2253.9+1609 & 343.49 & 16.15 & 478.95 & 14.7 & 0.84 & 457.94 & 17.66 & 0.82 \\ 
J2311.0+3425 & 347.77 & 34.42 & 430.92 & 96.92 & 0.13 & 493.96 & 197.77 & 0.21 \\ 
J2321.9+3204 & 350.47 & 32.07 & 10.63 & 252.95 & 0.35 & 523.98 & 91.59 & 0.36 \\ 
J2321.9+2734 & 350.5 & 27.55 & 302.83 & 41.0 & 0.75 & 302.83 & 364.91 & 0.73 \\ 
J2338.0-0230 & 354.49 & -2.52 & 263.81 & 137.32 & 0.63 & 316.84 & 224.33 & 0.78 \\ 
J2348.0-1630 & 357.01 & -16.52 & 1277.5 & 0.72 & 0.48 & 1277.5 & 0.0 & 0.27 \\ \hline
\label{tab:delayed_corr_results}
\end{longtable}

\begin{figure*}
\centering

\gridline{
  \fig{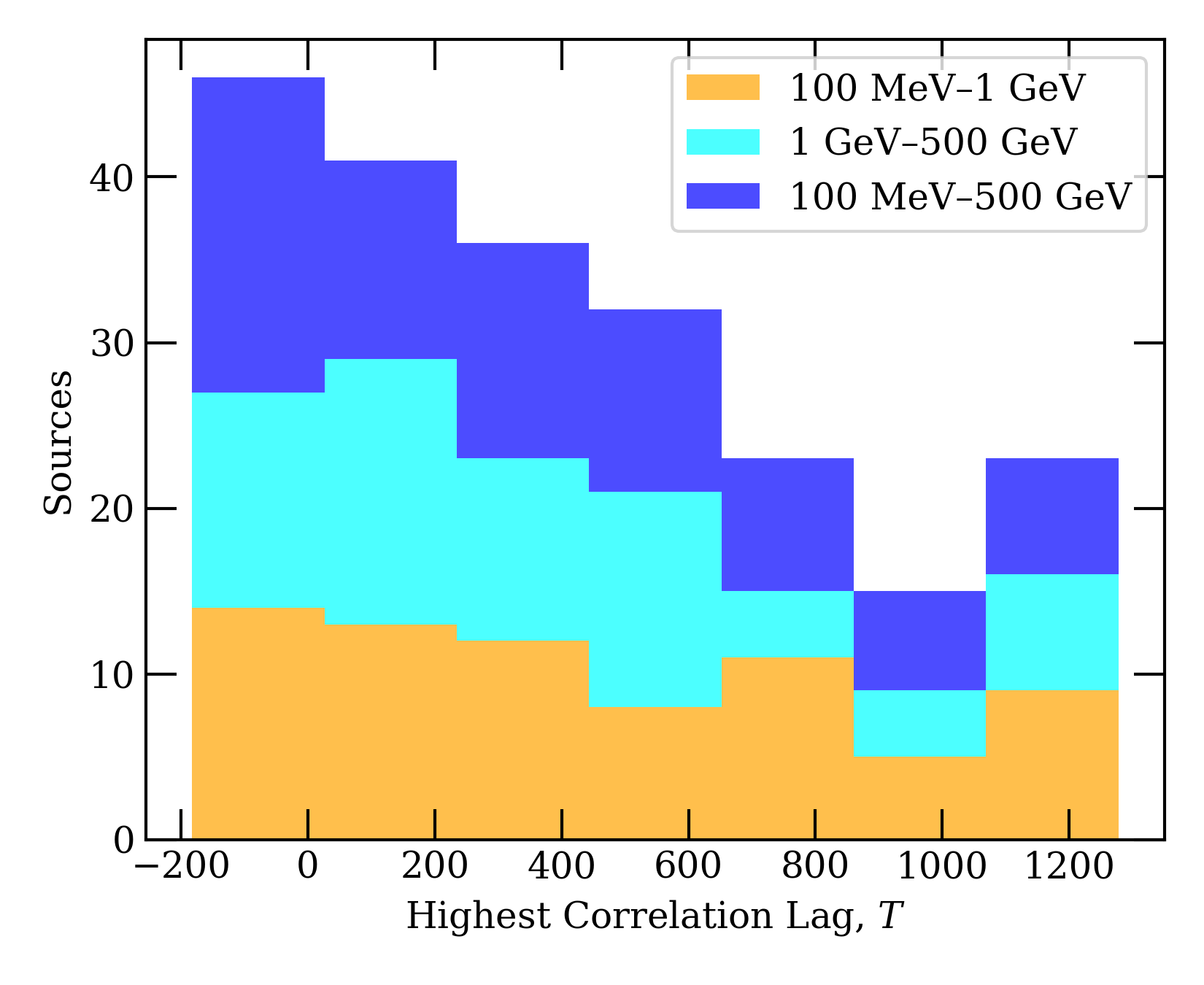}{0.5\textwidth}{}
  \fig{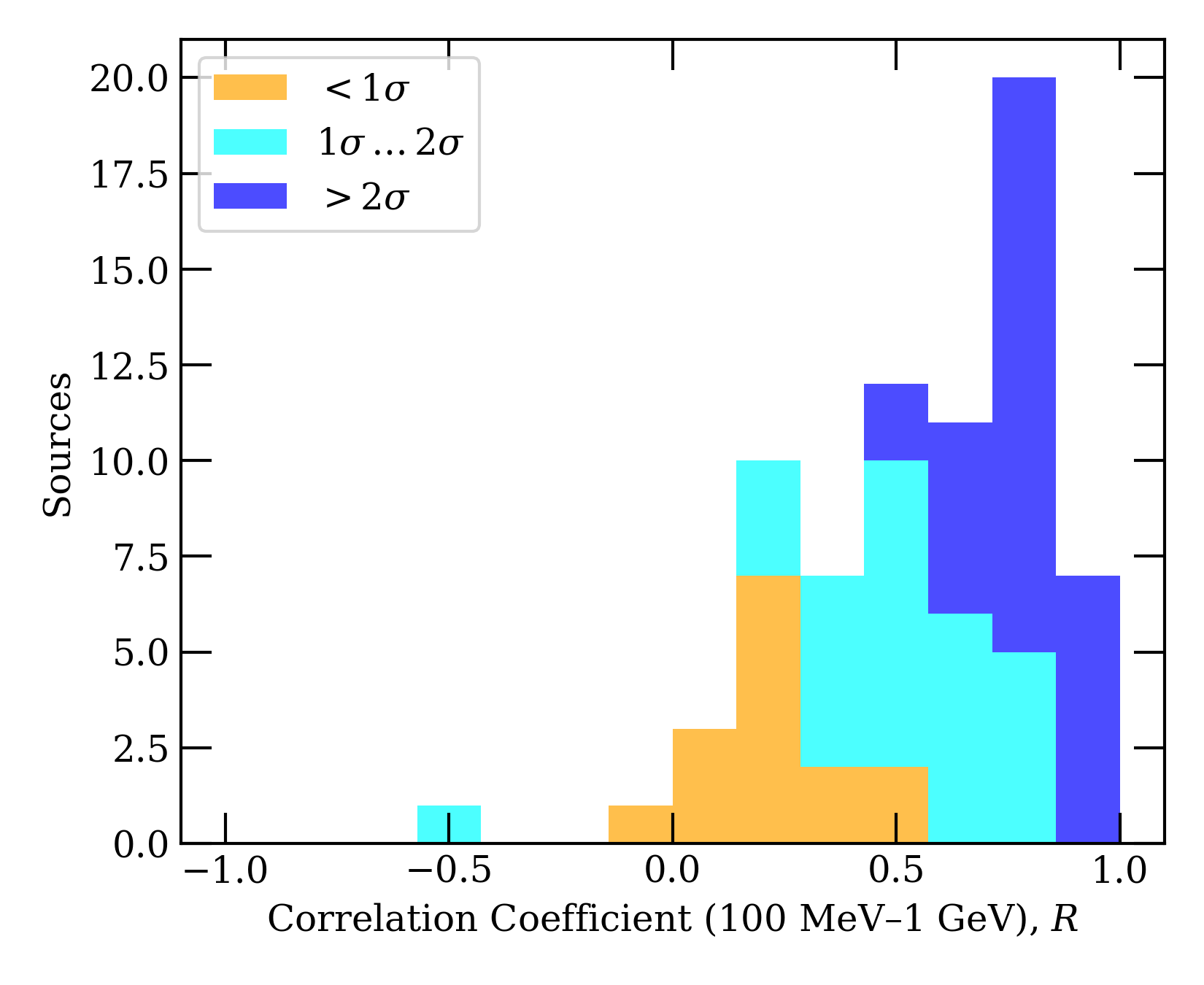}{0.5\textwidth}{}
}
\vspace{-3.0em}
\gridline{
  \fig{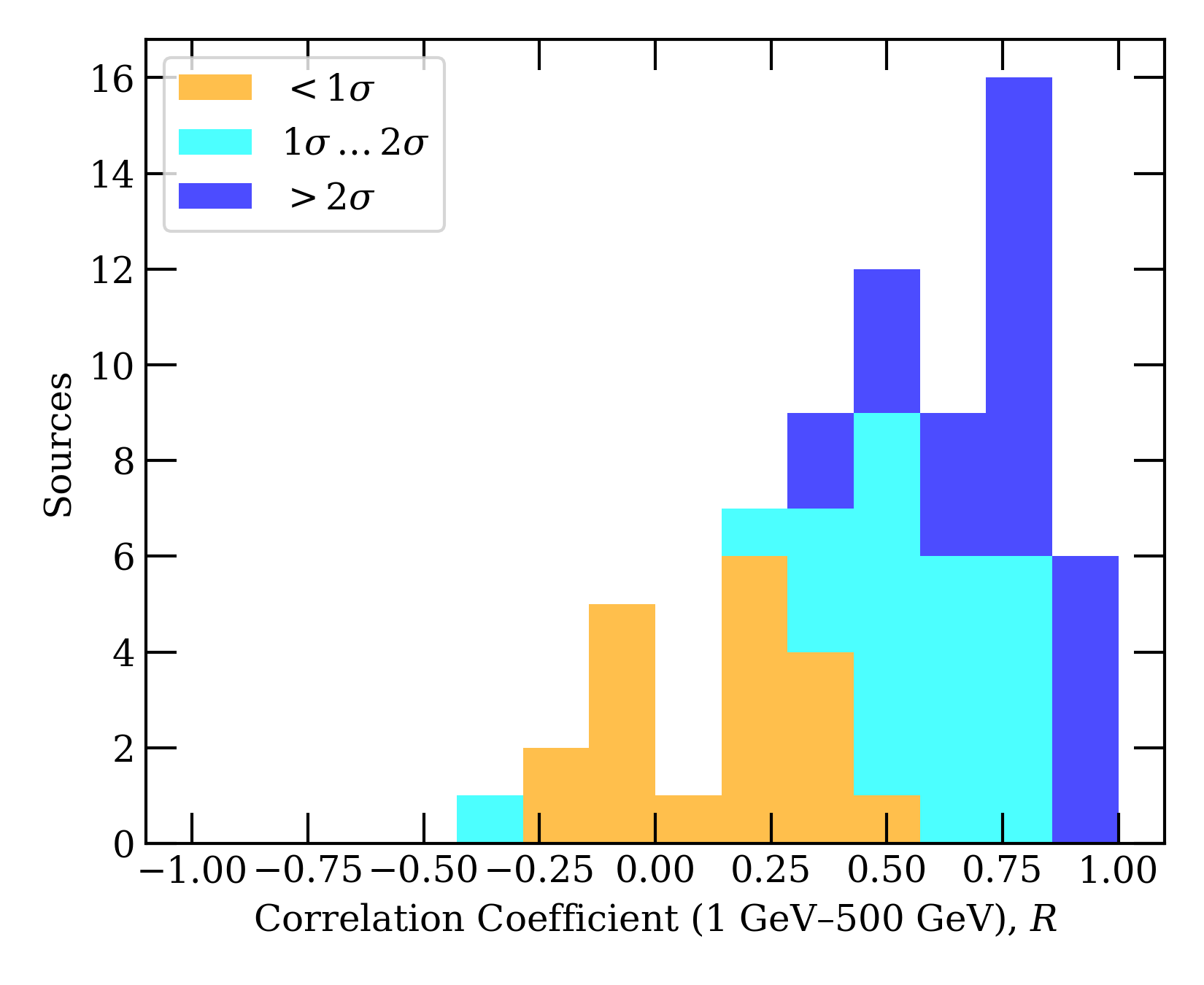}{0.5\textwidth}{}
  \fig{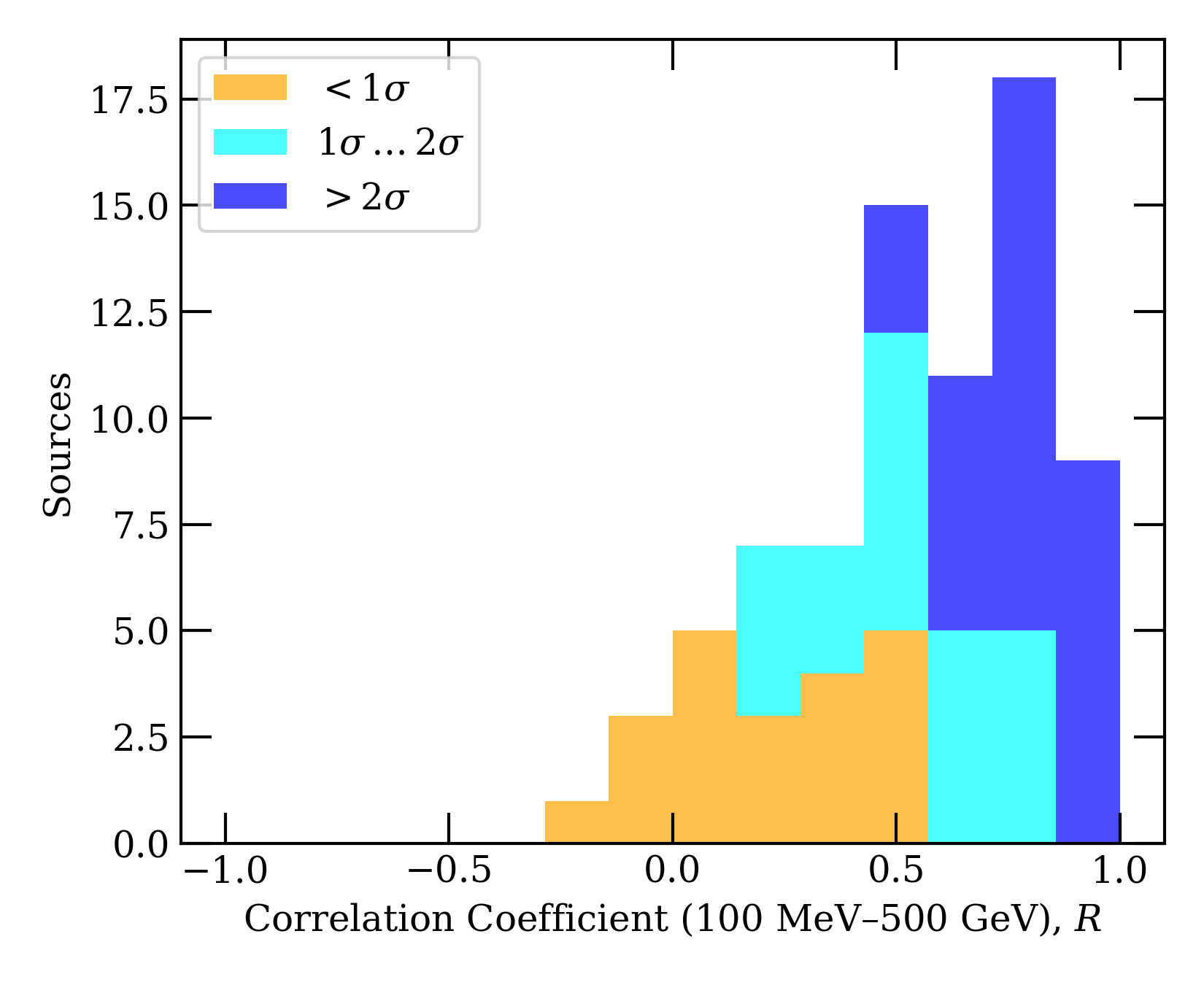}{0.5\textwidth}{}
}
\vspace{-2.0em}
\caption{Best-fit time lags and estimated significances for a delayed gamma-ray-radio flare. We find that only a small selection of sources shows some evidence for temporal correlation on this delayed timescale of 0.5–3.5 years. We plot these highest-correlation lags, $T_{\textrm{delay}}$, in the top left corner. Results for all gamma-ray energy bands are provided. We also provide distributions of estimated significances for each gamma-ray energy band. Here, significance is relative to the central 68$\%$ and 95$\%$ confidence containment regions constructed from randomized trials. }
\label{fig:bf_lags}

\end{figure*}

To compare the best-fit time lags, we provide a distribution of these values in Figure \ref{fig:bf_lags}. The significances for each gamma-ray band are also compared in the same figure. While many sources show preference for a correlation with minimal relative time lag (on the order of a few months) between radio and gamma-ray emission, about half of the population has a highest-correlation time lag within the range of 0.5 to 3.5 years.

\section{Stacked Correlations}

\begin{figure}
\centering

\includegraphics[width=0.8\linewidth]{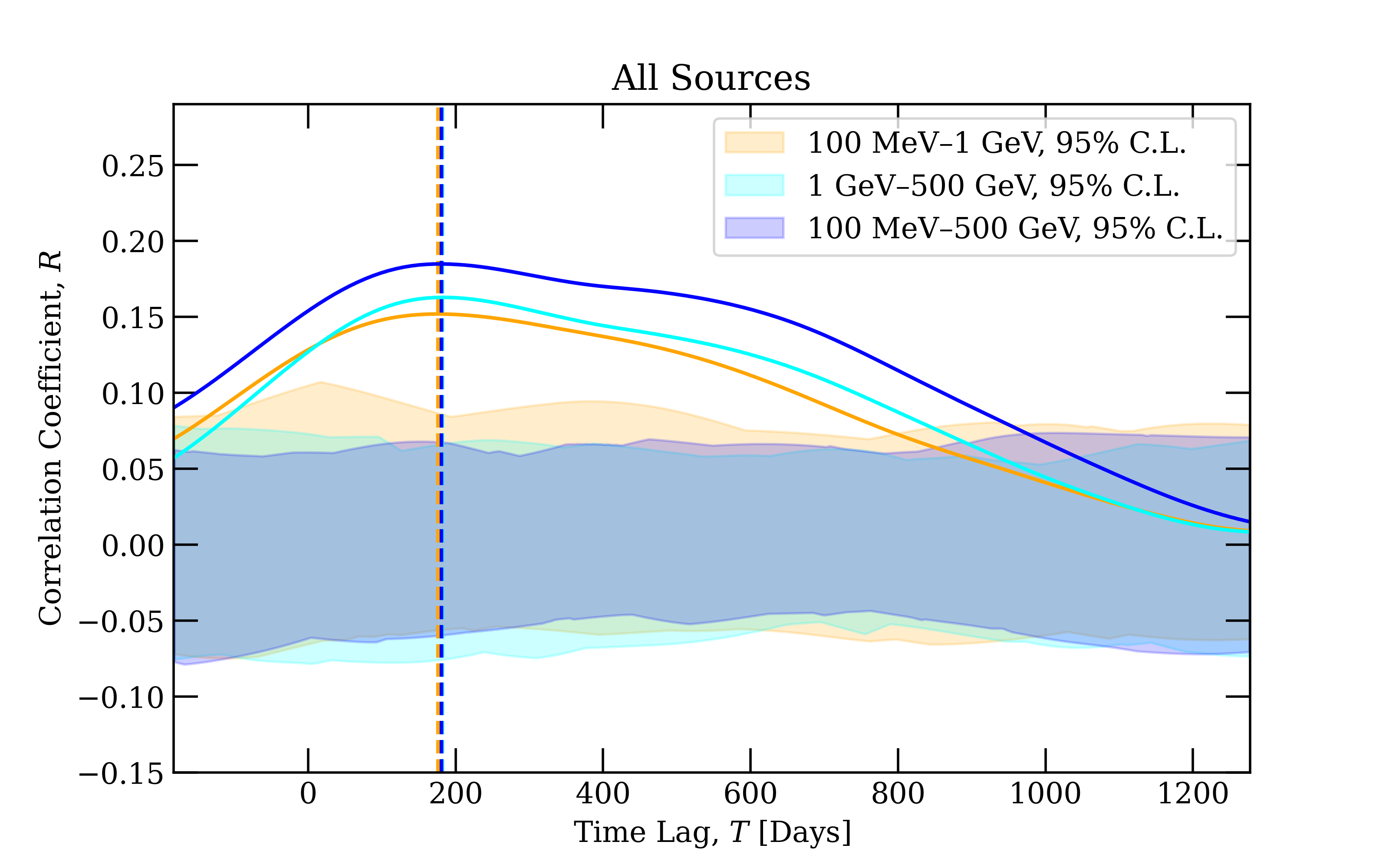}
\includegraphics[width=0.8\linewidth]{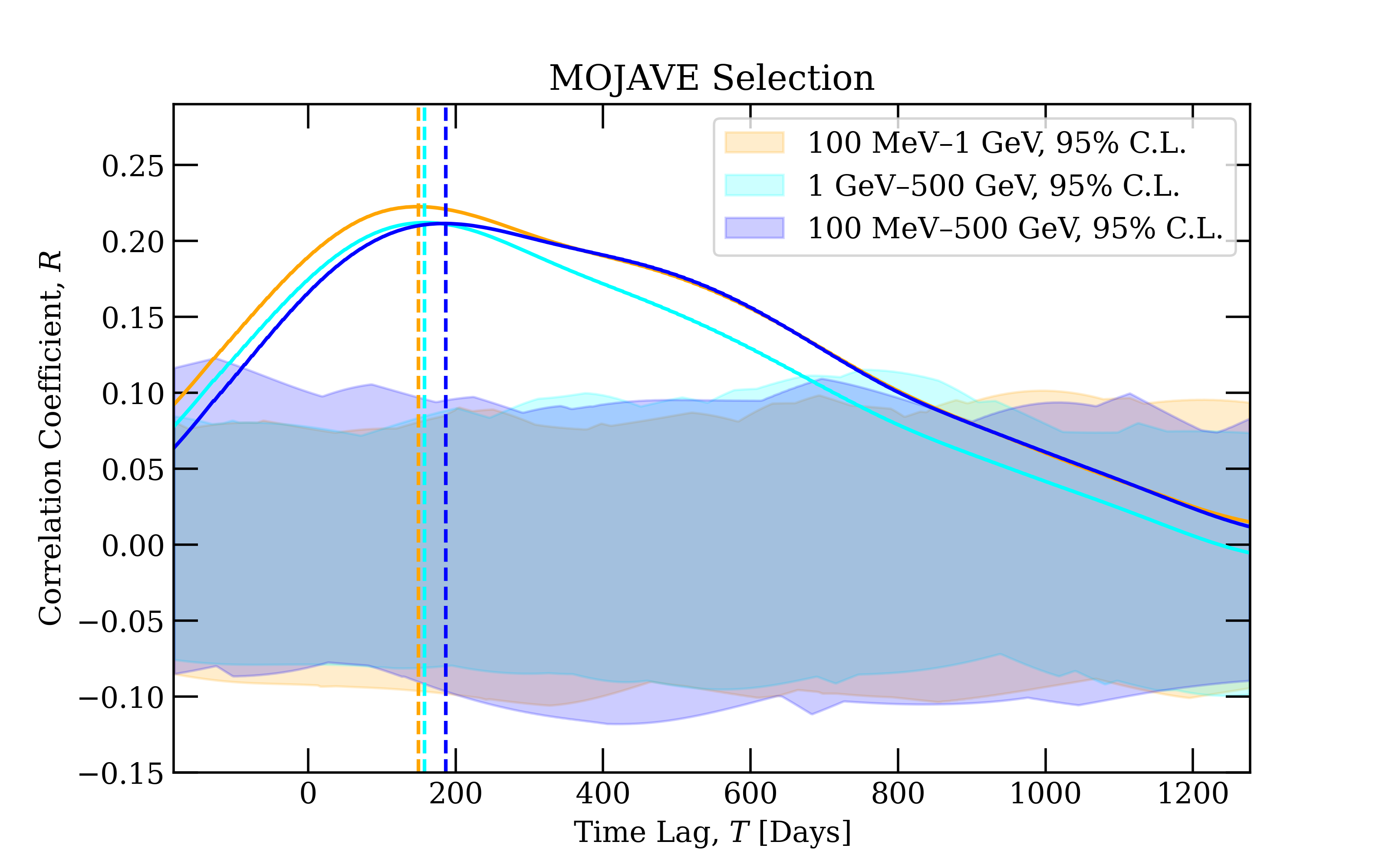}

\caption{Correlation coefficient as a function of time lag. Here we plot the correlation coefficient for the three gamma-ray energy ranges considered in this work. At the top, the correlation result for the set of all sources is shown. At the bottom, the result from the MOJAVE selection is plotted. The shaded bands indicate the 95$\%$ confidence regions determined for each light curve selection. The highest correlation time delays for all selections are indicated by dashed, vertical lines. These lines slightly overlap around the six month time. In all cases, a maximal correlation with significance beyond the 95$\%$ confidence level band is found.}
\label{fig:stack_corrs}
\end{figure}

We perform two sets of stacked correlations, considering both the selection of sources based on MOJAVE data and the set of all sources. As a majority of sources demonstrated a lag closer to zero within the individual source searches, we test this larger time window within the stacked correlation. Specifically, we have considered a range between -0.5 to 3.5 years. We provide the results of the stacked correlations for our different gamma-ray energy bands and selections in Table \ref{tab:stack_res}. We also plot the resulting correlation coefficient as a function of lag value for each case in Figure \ref{fig:stack_corrs}. 

\begin{table}[]
    \centering 
    \begin{tabular}{c|c|c|c|c|c} \hline 
    Selection & Gamma–Ray Energy Band & $T_{\textrm{stack}}$ [Days] &  
    $\Delta T_{\textrm{stack}}$ [Days] &  
    $R_{\textrm{stack}}$ & 
    $\sigma_{\textrm{stack}}$ \\ \hline 
    
    \multirow{3}{*}{All Sources} 
    & 100 MeV–1 GeV   & 175.75 & 24.71 & 0.15 & $> 2\sigma$ \\
    & 1 GeV–500 GeV   & 181.75 & 22.1  & 0.16 & $> 2\sigma$ \\
    & 100 MeV–500 GeV & 180.75 & 26.84 & 0.18 & $> 2\sigma$ \\ \hline 
    
    \multirow{3}{*}{MOJAVE Selection} 
    & 100 MeV–1 GeV   & 149.73 & 17.4 & 0.22 & $> 2\sigma$ \\
    & 1 GeV–500 GeV   & 157.73 & 14.41  & 0.21 & $> 2\sigma$ \\
    & 100 MeV–500 GeV & 186.75 & 18.44 & 0.21 & $> 2\sigma$ \\ \hline 
    \end{tabular}
    \caption{Correlation results for our selection of stacked correlations. Here, we provide the highest correlation time lag and uncertainty, $T_{\textrm{stack}}$ and $\Delta T_{\textrm{stack}}$, as well as the corresponding correlation coefficient, $R_{\textrm{stack}}$. The significance, $\sigma_{\textrm{stack}}$, is also provided for either result.   }
    \label{tab:stack_res}
\end{table}

In all stacked correlations, a significant result in excess of the 95$\%$ confidence limit band is found. All stacks show a highest-correlation time lag around six months. We also see a slight change in inflection of the correlation curve. This may suggest a population of sources with larger best-fit time lags or radio activity over an extended time period. The MOJAVE selection shows a slightly stronger correlation, potentially reflecting the more refined selection of jet core activity. 

\section{Conclusions}
In this work, we have searched for a correlation between RATAN-600 blazar radio data and Fermi-LAT light curves constructed for several energy bands. We examined both $\sim$80 individual sources as well as stacked populations. While previous studies have indicated a general delay between gamma-ray and radio data on the scale of a few months, we have considered a larger relative search window. This was partially motivated by the flaring phenomenology of likely neutrino-bright blazar sources, as well as the general dynamics and structure of the gamma-ray-radio jet. 

While a number of sources exhibit evidence for a highest-correlation lag within -0.5 to 0.5 years, more than half of our tested population was associated with time delays between 0.5 to 3.5 years. This may be due to the longer duration of the period, however we also found that the best-fit time delay for all stacked searches was on the order of six months. These stacked results were significant beyond a 2$\sigma$ significance. As our background trials were constructed from mismatched light curves, it appears that there is a substantial, significant correlation at these higher time delay values. While sources with best-fit time lags closer to zero contribute, it's likely that a significant signal is also driven by radio flaring at later times. The radio activity of a selection of blazars follows or continues after gamma-ray activity, potentially related to either slow jet expansion and propagation or to a more complex jet structure.

We considered three gamma-ray energy bands in our analysis, 100 MeV–1 GeV, 1 GeV–500 GeV, and the integral band between 100 MeV and 500 GeV. We have found the highest significance stacked result using this integral band, and comparing the maximal correlation coefficient, $R$, and the 95$\%$ confidence region. This may be due to the higher statistical significance of this gamma-ray data used for light curve construction. The two distinct energy bands were constructed to test for potential lag differences owing to preferential gamma-ray absorption by the broad line region. We did not find any statistically significant difference between these two energy bands in our stacked analyses. 

While we have demonstrated some evidence for a potential population of blazars with longer gamma-ray-radio delays or extended periods of radio emission, further work may focus on isolating and explaining these sources. In particular, more refined models of the central jet structure and evolution may help to explain the relation between multimessenger observations. 

\section*{Acknowledgments}

E.K. thanks funding from the NKFIH excellence grant TKP2021-NKTA-64. A.K. thanks support from NSF grant PHY-2237581. This paper makes use of publicly available \textit{Fermi}-LAT data provided online by the \url{https://fermi.gsfc.nasa.gov/ssc/data/access/} Fermi Science Support Center. This research has made use of data from the MOJAVE database that is maintained by the MOJAVE team \citep{2018ApJS..234...12L}. The National Radio Astronomy Observatory is a facility of the National Science Foundation operated under the cooperative agreement by Associated Universities, Inc.  We acknowledge the use of data from the Astrogeo Center database maintained by Leonid Petrov. This work was supported in part through computational resources and services provided by the Institute for Cyber-Enabled Research at Michigan State University.

\nocite{*}
\bibliographystyle{aasjournal}
\bibliography{notes}

\section{Appendix}

\begin{longtable}{lcccccc}
\caption{Correlation results for a radio-delayed flare. Given our search window of -0.5 to 3.5 years, we provide the tested lag and uncertainty in days, $T_{\textrm{delay}}$ and $\Delta T_{\textrm{delay}}$, corresponding to the highest correlation coefficient $R_{\textrm{delay}}$. Results for the integral gamma-ray energy band are provided.} \\
\hline
 Source Name & R.A. [$\degree$] & Dec. [$\degree$] & $T_{\textrm{delay, integral}}$ &  $\Delta T_{\textrm{delay, integral}}$ &  $R_{\textrm{delay, integral}}$   \\ \hline
\endfirsthead

\hline 
Source Name & R.A. [$\degree$] & Dec. [$\degree$] & $T_{\textrm{delay, integral}}$ &  $\Delta T_{\textrm{delay, integral}}$ &  $R_{\textrm{delay, integral}}$   \\ \hline
\endhead

J0013.6-0424 & 3.48 & -4.4 & 1277.5 & 571.94 & -0.14 \\ 
J0018.8+2611 & 4.91 & 26.05 & 665.08 & 228.0 & 0.93 \\ 
J0112.8+3208 & 18.21 & 32.14 & 620.05 & 165.14 & 0.18 \\ 
J0118.9-2141 & 19.74 & -21.69 & 599.04 & 13.92 & 0.72 \\ 
J0132.7-1654 & 23.18 & -16.91 & 534.99 & 15.49 & 0.77 \\ 
J0144.6+2705 & 26.14 & 27.08 & 217.77 & 37.58 & 0.46 \\ 
J0152.2+2206 & 28.07 & 22.12 & -182.5 & 43.96 & 0.41 \\ 
J0238.6+1637 & 39.66 & 16.62 & -0.38 & 9.03 & 0.78 \\ 
J0239.7+0415 & 39.96 & 4.27 & 712.11 & 236.63 & 0.8 \\ 
J0312.8+0134 & 48.18 & 1.55 & 24.64 & 400.08 & 0.8 \\ 
J0343.4+3621 & 55.87 & 36.37 & 1272.5 & 195.9 & 0.79 \\ 
J0348.6-2749 & 57.16 & -27.82 & 578.02 & 26.56 & 0.95 \\ 
J0405.6-1308 & 61.39 & -13.14 & 180.75 & 39.69 & 0.75 \\ 
J0416.5-1852 & 64.15 & -18.85 & 1.63 & 546.22 & 0.78 \\ 
J0423.3-0120 & 65.81 & -1.34 & 248.8 & 217.02 & 0.76 \\ 
J0423.9+4150 & 65.98 & 41.83 & 1047.34 & 534.18 & 0.57 \\ 
J0434.1-2014 & 68.53 & -20.25 & -182.5 & 532.5 & 0.46 \\ 
J0442.6-0017 & 70.66 & -0.3 & -182.5 & 369.28 & 0.46 \\ 
J0449.1+1121 & 72.28 & 11.36 & 30.65 & 17.61 & 0.89 \\ 
J0453.1-2806 & 73.31 & -28.13 & 183.75 & 314.99 & 0.32 \\ 
J0457.0-2324 & 74.26 & -23.41 & 104.7 & 18.46 & 0.65 \\ 
J0502.4+0609 & 75.56 & 6.15 & 741.13 & 9.37 & 0.66 \\ 
J0510.0+1800 & 77.51 & 18.01 & -62.42 & 8.9 & 0.56 \\ 
J0532.6+0732 & 83.16 & 7.55 & 1149.41 & 489.88 & 0.56 \\ 
J0533.3+4823 & 83.31 & 48.38 & 380.89 & 26.54 & 0.88 \\ 
J0539.9-2839 & 84.97 & -28.67 & 293.83 & 29.32 & 0.44 \\ 
J0654.4+4514 & 103.6 & 45.24 & -33.4 & 351.45 & 0.6 \\ 
J0656.3-0322 & 104.05 & -3.38 & -158.48 & 455.78 & 0.24 \\ 
J0659.6-2742 & 104.95 & -27.75 & 297.83 & 172.81 & 0.63 \\ 
J0718.0+4536 & 109.46 & 45.63 & 749.14 & 86.81 & 0.06 \\ 
J0725.2+1425 & 111.32 & 14.42 & 645.07 & 51.53 & 0.29 \\ 
J0748.6+2400 & 117.15 & 24.01 & 432.92 & 115.11 & 0.44 \\ 
J0754.4-1148 & 118.61 & -11.79 & -43.4 & 158.05 & 0.39 \\ 
J0808.2-0751 & 122.06 & -7.85 & 78.68 & 29.87 & 0.85 \\ 
J0809.3+4053 & 122.23 & 40.88 & 504.97 & 312.2 & -0.16 \\ 
J0829.0+1755 & 127.27 & 17.9 & 245.79 & 371.1 & 0.14 \\ 
J0830.8+2410 & 127.72 & 24.18 & 274.81 & 80.87 & 0.56 \\ 
J0850.0+5108 & 132.49 & 51.14 & -53.41 & 191.74 & 0.04 \\ 
J0920.9+4441 & 140.24 & 44.7 & 18.64 & 34.25 & 0.83 \\ 
J0921.6+6216 & 140.4 & 62.26 & 18.64 & 141.05 & 0.63 \\ 
J0923.5+4125 & 140.88 & 41.42 & 102.7 & 28.23 & -0.14 \\ 
J1006.7-2159 & 151.69 & -21.99 & 584.03 & 254.32 & 0.26 \\ 
J1114.5-0819 & 168.63 & -8.28 & -182.5 & 512.73 & 0.48 \\ 
J1118.6-1235 & 169.57 & -12.55 & 801.17 & 131.47 & 0.91 \\ 
J1127.0-1857 & 171.77 & -18.95 & 1029.33 & 10.24 & 0.93 \\ 
J1129.8-1447 & 172.53 & -14.82 & 939.27 & 108.72 & 0.79 \\ 
J1135.7-0427 & 173.99 & -4.47 & -41.4 & 289.08 & 0.26 \\ 
J1345.8+0706 & 206.45 & 7.11 & 978.3 & 394.12 & 0.28 \\ 
J1357.1+1921 & 209.27 & 19.32 & 669.08 & 459.14 & 0.5 \\ 
J1613.6+3411 & 243.42 & 34.21 & 702.11 & 71.3 & 0.88 \\ 
J1616.7+4107 & 244.27 & 41.11 & 99.69 & 450.66 & 0.88 \\ 
J1631.2+4926 & 247.82 & 49.46 & 1008.32 & 402.92 & 0.8 \\ 
J1635.2+3808 & 248.81 & 38.13 & 86.68 & 6.95 & 0.73 \\ 
J1642.9+3948 & 250.74 & 39.81 & -96.44 & 30.09 & 0.7 \\ 
J1728.4+0427 & 262.1 & 4.45 & -80.43 & 577.86 & 0.71 \\ 
J1733.0-1305 & 263.26 & -13.08 & 1277.5 & 0.0 & 0.42 \\ 
J1734.3+3858 & 263.58 & 38.96 & 669.08 & 180.87 & 0.42 \\ 
J1740.5+5211 & 265.15 & 52.2 & 266.81 & 16.96 & 0.69 \\ 
J1753.7+2847 & 268.42 & 28.8 & 871.22 & 47.97 & 0.96 \\ 
J2136.2+0032 & 324.16 & 0.7 & 1277.5 & 84.67 & 0.48 \\ 
J2146.4-1528 & 326.59 & -15.43 & 1179.43 & 155.69 & 0.72 \\ 
J2147.1+0931 & 326.79 & 9.5 & 101.69 & 416.27 & 0.2 \\ 
J2158.1-1501 & 329.52 & -15.02 & 371.88 & 255.48 & 0.11 \\ 
J2219.2-0342 & 334.72 & -3.59 & 389.89 & 100.52 & 0.59 \\ 
J2219.2+1806 & 334.81 & 18.11 & -19.39 & 133.58 & 0.56 \\ 
J2225.6+2120 & 336.41 & 21.3 & 333.85 & 384.26 & 0.13 \\ 
J2229.7-0832 & 337.42 & -8.55 & -2.38 & 13.34 & 0.82 \\ 
J2232.6+1143 & 338.15 & 11.73 & 37.65 & 103.38 & -0.09 \\ 
J2236.3+2828 & 339.09 & 28.48 & 71.67 & 17.01 & 0.55 \\ 
J2243.9+2021 & 340.97 & 20.35 & 511.98 & 218.81 & 0.59 \\ 
J2253.9+1609 & 343.49 & 16.15 & 478.95 & 19.97 & 0.84 \\ 
J2311.0+3425 & 347.77 & 34.42 & 451.93 & 93.66 & 0.16 \\ 
J2321.9+3204 & 350.47 & 32.07 & 523.98 & 164.7 & 0.36 \\ 
J2321.9+2734 & 350.5 & 27.55 & 302.83 & 430.68 & 0.74 \\ 
J2338.0-0230 & 354.49 & -2.52 & 242.79 & 65.06 & 0.61 \\ 
J2348.0-1630 & 357.01 & -16.52 & 1277.5 & 0.0 & 0.44  
\\ \hline 
\label{tab:delayed_corr_results_int}
\end{longtable}

\end{document}